\documentstyle[11pt,fleqn]{article}
\begin{document}

\begin{titlepage}

\begin{flushright}
Freiburg--THEP 95/18\\
Juli 1995
\end{flushright}
\vspace{1.5cm}

\begin{center}
\large\bf
{\LARGE\bf An axially-symmetric Newtonian Boson Star}\\[1cm]
\rm
{B. Schupp}\\
{and}\\
{J.J. van der Bij}\\[.5cm]

{\em Albert--Ludwigs--Universit\"{a}t Freiburg,
           Fakult\"{a}t f\"{u}r Physik}\\
      {\em Hermann--Herder Str.3, D-79104 Freiburg, Germany}\\[1.5cm]

\end{center}
\normalsize

\begin{abstract}

A new solution to the coupled gravitational and scalar field equations for
a condensed boson field is found in Newtonian approximation. The solution
is axially symmetric, but not spherically symmetric. For N particles the mass
of the object is given by $M = Nm - 0.02298 N^3 G_N^2 m^5$, to be compared with
$M = Nm  - 0.05426 N^3 G_N^2 m^5$ for the spherically symmetric case.

\end{abstract}
\end{titlepage}

Recent developments in particle physics and cosmology suggest
that evolving scalar fields may have played an important role
in the evolution of the early universe, for instance in primordial
phase transitions, and that they may make up the missing dark matter.
Models for galaxy formation using cold dark matter and the inflationary
scenario suggest that the ratio of baryonic (luminous) matter to dark
matter can be of the order of 10 \%. These facts naturally raise
the question whether cold gravitational equilibrium configurations
of massive scalar fields - Bose stars - may exist and whether such
configurations are dynamically stable. For bosonic fields interacting
only via gravity spherically symmetric equilibrium solutions
were studied by Kaup~[1] as well as Ruffini and Bonazzola~[2]
by solving the coupled Einstein-Klein-Gordon equations.
They analyzed only the zero-node solutions, corresponding to the
lowest energy state. The results of ref.[1,2] have been confirmed and
extended later on. For reviews we refer to ref.[3,4]. Recently the
suggestion has been made, that the halo of galaxies is itself a
condensed bosonic object[5,6]. This model was studied in the Newtonian
approximation, where reasonable agreement with experimental rotation
curves was found. All studies sofar have been restricted to non-rotating
objects, i.e. spherically symmetric solutions. In this letter we make a
first approach in studying axially symmetric solutions. For simplicity we
restrict ourselves here to the Newtonian approximation.

The Newtonian treatment of self-gravitating bosons of mass m interacting
only gravitationally has been studied in ref.[2,7]. For condensed bosons
at T=0 the equations reduce to two coupled equations for the gravitational
potential and the Schr\"odinger field. The gravitational potential V
satisfies the Poisson equation $$\Delta V = 4\pi G_N \rho \eqno(1)$$
where the mass density is given by $$\rho = Nm \psi^* \psi \eqno(2)$$ The one
particle wave function $\psi$ is determined by the Schr\"odinger equation
$$ - \Delta \psi + 2m(E+mV)
\psi = 0 \eqno(3)$$ together with the normalization $$\int d^3r \psi^*\psi = 1
\eqno(4)$$
After rescaling $$\hat x = 2 m^3 G_N x \eqno(5)$$ $$\phi = \sqrt{4 \pi}
(2 m^3 G_N N)^{-3/2} \psi \eqno(6)$$ $$\hat V = (2 G_N^2 N^2 m^4)^{-1} V
\eqno(7)$$
$$\hat E = (2 G_N^2 N^2 m^5)^{-1}E \eqno(8)$$ the system of equations reduces
to
the simple form $$\Delta \hat V =  \phi^2 \eqno(9)$$ and $$\Delta \phi - \hat V
\phi =
\hat E \phi \eqno(10)$$ The search for a solution is then reduced to looking
for an eigenfunction with the correct norm $$\int_0^{\infty} \phi_{\hat E}^2
d^3r =4 \pi \eqno(11)$$ In the case of a spherically symmetric solution the
equations can be reduced to a set of ordinary differential equations. These can
be solved relatively simply. In our case we are also interested in solutions,
that
have only an axial symmetry. Here the system of equations cannot be reduced to
ordinary differential equations. To solve the resulting equations, we used the
method of finite elements [8,9]. In practice the following procedure was
adopted.
The equations were rewritten in cylindrical coordinates. As boundary condition
we took $$\phi (r,z)=0 \hspace{5pt} {\rm for} \hspace{5pt} r^2+z^2 =R 
\eqno(12)$$ When
R is large enough this gives a sufficient approximation for the condition $\phi
\rightarrow 0$ at infinity.
Furthermore there is the condition $$\partial_r \phi (r=0,z) = 0 \eqno(13)$$
It is to be noticed that the equations are invariant under the parity
transformation $z\rightarrow -z$, $\hat V \rightarrow \hat V$, $\phi
\rightarrow \pm \phi$. The solutions therefore fall into positive and negative
parity classes. Using this symmetry it is therefore sufficient to  solve the
equations in a quadrant in the r,z plane with the following boundary
conditions:\\
Dirichlet for $\phi $ and $\hat V$ at $r^2+z^2=R$,\\
Neumann for $\phi$ and $\hat V$ at $r=0$ and for $\hat V$ at $z=0$.\\
For positive parity one has then
the Neumann condition for $\phi$ at $z=0$, while for negative parity one has
the Dirichlet condition. With these conditions the problem is  well
defined in a finite domain and can be solved with the finite element method.
Because the equations are partial differential equations one can typically find
this way the lowest energy solutions corresponding to the given boundary
conditions. The solution for positive parity is given in the figures
1 and 2 . The corresponding eigenvalue is given by $\hat E = 0.081385$, which
is in perfect agreement with the results in the literature and the integration
after imposing spherical symmetry. This good agreement proves that the finite
element method can be satisfactorily applied to this problem. The corresponding
results for the negative parity solution is given in figures 3 and 4. We found
an energy value of $\hat E = 0.034465$. These numbers correspond to the
following mass formulae,  $$M = Nm  - 0.05426 N^3 G_N^2 m^5 \eqno(14)$$ for the
spherically symmetric case and $$M = Nm - 0.02298 N^3 G_N^2 m^5 \eqno(15)$$ for
the
axially symmetric  case. The maximum masses and particle numbers are given by
$$M=1.6524 m_{Pl}^2/m \hspace{5pt} and  \hspace{5pt}
N=2.4786 m_{Pl}^2/m^2 \eqno(16)$$
respectively
$$M=2.5391 m_{Pl}^2/m \hspace{5pt} and \hspace{5pt} N=3.8086 m_{Pl}^2/m^2 
\eqno(17)$$\\

{\bf References}
\begin{enumerate}
\item D.~J.~Kaup, Phys. Rev. 172 (1968), 1331.
\item R.~Ruffini and S.~Bonazzola, Phys. Rev. 187 (1969), 1767.
\item P.~Jetzer, Phys. Rept. 220 (1992), 163.
\item T.~D.~Lee and Y.~Pang, Phys. Rept. 222 (1992), 251.
\item H.~Dehnen and B.~Rose, Astrophys. and Space Science, 207 (1993), 133.
\item Sang-Jin Sin, Phys. Rev. D50 (1994), 3650.
\item M.~Membrado, A.~F.~Pacheco and J.~Sa\~nudo, Phys. Rev. A39 (1989), 4207.
\item P.~G.~Ciarlet, The finite elements method for elliptic problems,
North Holland (1978).
\item H.~R.~Schwarz, Methode der finiten Elemente, B.~G.~Teubner (1991).
\end{enumerate}

{\bf Figure Captions}
\begin{enumerate}
\item Presented is the behaviour of the funktion $\phi (r,z)$ for the
spherically symmetric case. The straight line corresponds to the z-axis.
The semi-circle corresponds to $r^2+z^2=50$. Outside the plotted area
$\phi \approx 0$.
\item Presented is the behaviour of the gravitational potential.
Outside the plotted region one has asymptotically $\hat V = (r^2+z^2)^{-1/2}$
\item The same as fig.(1), but for the axially-symmetric case. Here the
semi-circle is at R=100.
\item The gravitational potential for the axially-symmetric case.
\end{enumerate}
\end{document}